\documentclass[journal=jpc,manuscript=article]{achemso}
\setkeys{acs}{articletitle=true}

\usepackage{chemformula} 
\usepackage[T1]{fontenc} 
\usepackage{xcolor}
\usepackage{longtable}
\usepackage{tablefootnote}
\usepackage{amsmath} 
\usepackage{amssymb}
\usepackage{tcolorbox} 
\usepackage{enumerate} 
\usepackage{mathtools}
\usepackage{multirow}	

\setkeys{acs}{
etalmode = truncate,
maxauthors=10
}

\newcommand{\onlinecite}[1]{\hspace{-2 ex} \citenum{#1}}
\newcommand{\wn}{~\rm{cm}^{-1}}

\author{Tianxiang Chen}
\affiliation{Department of Chemistry, The Johns Hopkins University, Baltimore, MD 21218, USA}
\author{Chaoqun Zhang}
\affiliation{Department of Chemistry, The Johns Hopkins University, Baltimore, MD 21218, USA}
\author{Lan Cheng}
\email{lcheng24@jhu.edu}
\affiliation{Department of Chemistry, The Johns Hopkins University, Baltimore, MD 21218, USA}
\author{Kia Boon Ng}
\email{kbng@triumf.ca}
\affiliation{TRIUMF, 4004 Wesbrook Mall, Vancouver, BC V6T 2A3, Canada}
\author{Stephan Malbrunot-Ettenauer}
\affiliation{TRIUMF, 4004 Wesbrook Mall, Vancouver, BC V6T 2A3, Canada}
\alsoaffiliation{Department of Physics, University of Toronto, Toronto M5S 1A7, Canada}
\author{Victor V. Flambaum}
\affiliation{School of Physics, University of New South Wales, Sydney 2052, Australia}
\author{Zack Lasner}
\affiliation{Department of Physics, Harvard University, Cambridge, Massachusetts 02138, USA}
\alsoaffiliation{Harvard-MIT Center for Ultracold Atoms, Cambridge, Massachusetts 02138, USA}
\email{zlasner@g.harvard.edu}
\author{John M. Doyle}
\affiliation{Department of Physics, Harvard University, Cambridge, Massachusetts 02138, USA}
\alsoaffiliation{Harvard-MIT Center for Ultracold Atoms, Cambridge, Massachusetts 02138, USA}
\author{Phelan Yu}
\affiliation{Division of Physics, Mathematics, and Astronomy, California Institute of Technology, Pasadena, CA 91125, USA}
\author{Chandler J. Conn}
\affiliation{Division of Physics, Mathematics, and Astronomy, California Institute of Technology, Pasadena, CA 91125, USA}
\author{Chi Zhang}
\affiliation{Division of Physics, Mathematics, and Astronomy, California Institute of Technology, Pasadena, CA 91125, USA}
\author{Nicholas R. Hutzler}
\affiliation{Division of Physics, Mathematics, and Astronomy, California Institute of Technology, Pasadena, CA 91125, USA}
\email{hutzler@caltech.edu}
\author{Andrew M. Jayich}
\affiliation{Department of Physics, University of California, Santa Barbara, California 93106, USA}
\author{Benjamin Augenbraun}
\affiliation{Department of Chemistry, Williams College, 47 Lab Campus Drive Williamstown, MA 01267, USA}
\author{David DeMille}
\affiliation{Department of Physics, University of Chicago, Chicago, Illinois 60637, USA}

\title
{Relativistic Exact Two-Component Coupled-Cluster Study of 
Molecular Sensitivity Factors for Nuclear Schiff Moments}

\begin{document}


\newpage
\begin{abstract}
  Relativistic exact two-component coupled-cluster calculations of 
  molecular sensitivity factors for nuclear Schiff moments (NSMs) are reported. We focus on molecules containing heavy nuclei, 
  especially octupole-deformed nuclei.
  Analytic relativistic coupled-cluster gradient techniques are used and serve as useful tools for identifying candidate molecules 
  that sensitively probe for physics beyond the Standard Model in the hadronic sector. Notably, these tools enable straightforward ``black-box'' calculations.
    Two competing chemical mechanisms that contribute to the NSM are analyzed, illuminating the physics of 
 ligand effects on NSM sensitivity factors. 

\end{abstract}


\section{Introduction}

Many open questions in fundamental physics, such as the Baryon Asymmetry~\cite{Dine2003BAU} and the Strong CP Problem~\cite{Kim2010StrongCP}, can be studied using precision measurements of fundamental symmetry violations in atoms and molecules~\cite{Safronova18,Hutzler2020PolyReview,Alarcon22}.  Molecules, in particular, have extreme internal electromagnetic environments which amplify the effects of symmetry-violating electromagnetic moments such as electric dipole moments (EDMs), nuclear magnetic quadrupole moments (MQMs), and nuclear Schiff moments (NSMs) -- all of which violate both $T-$ and $P-$ symmetries, and are sensitive to symmetry-violating physics beyond the Standard Model~\cite{Engel2013EDM}.  Sensitivity to nuclear symmetry violations via NSMs can be enhanced by around a thousand-fold in heavy nuclei possessing an octupole ($\beta_3$) deformation and non-zero nuclear spin~\cite{Arrowsmithkron23}, including statically deformed isotopes of radioactive species Fr, Ra, Ac, Th, and Pa, as well as dynamically deformed isotopes of stable species Eu and Dy, all of which present both experimental and theoretical challenges. 

Experiments to search for NSMs using atoms and molecules rely on the symmetry-violating interaction between the nuclear spin and the electronic orbitals by precise comparisons of the energy of two states with opposite nuclear spin orientation~\cite{Safronova18}.  For a given NSM, the induced energy shift, which is the experimental observable, depends on details of the electronic structure and how the electrons interact with the nucleus. 
Specifically, the induced energy shift is proportional to a
NSM sensitivity factor of the molecular state used in the search. \cite{Safronova18}
In order to both design and interpret experiments, one needs to calculate this NSM molecular sensitivity parameter by performing quantum-chemical calculations.


Relativistic {{Hartree-Fock}} and density functional theory, \cite{Laerdahl97,Gaul19,Gaul20,Gaul20a,Gaul24} coupled-cluster, \cite{Petrov02,Kudashov13,Skripnikov20,Abe20,Gaul24} and mulitreference configuration interaction (MRCI) \cite{Hubert22,Marc23} calculations of NSM molecular sensitivity factors have been reported.
To improve the robustness and efficiency for calculations with high-level treatments of electron-correlation effects,
in this work we extend analytic relativistic coupled-cluster gradient techniques to calculations of NSM molecular sensitivity factors. Molecular structure calculations relevant to experimental searches for NSMs in heavy nuclei, including the exceptionally important case of octupole
deformed nuclei, are reported. Further, we analyze two competing chemical mechanisms contributing to the NSM molecular sensitivity parameters and discuss the implication to engineering suitable candidate molecules. 

In the following, we first provide general background information about relativistic coupled-cluster theory 
and precision spectroscopic searches for time-reversal symmetry violation. 
We then summarize theory and computational details, and present discussions of the computational results. 
Finally, we provide a summary and an outlook.



 
 \section{Generalities}

 \subsection{Coupled-Cluster Theory}
 
Coupled-cluster (CC) methods \cite{Bartlett07,Crawford00} provide size-extensive 
and systematically improvable treatments of electron correlation in atoms and molecules.
The CC singles and doubles (CCSD) \cite{Purvis82} method has been shown to capture the majority of electron-correlation effects
in electronic states dominated by a single Slater determinant.
The inclusion of triple and quadruple excitations 
in the CC singles doubles triples (CCSDT) \cite{Noga87a,Scuseria88,Watts90} method and the
CC singles doubles triples quadruples (CCSDTQ) \cite{Oliphant91,Kucharski92,Kallay01,Hirata03} method
paves the way to essentially quantitative treatments of electron correlation. 
More approximate variants, including the CCn models \cite{Christiansen95a,Koch97} and 
a variety of non-iterative treatments of triple and quadruple excitations \cite{Raghavachari89,Bartlett90,Stanton97,Crawford98,Kucharski98,Hirata01,Bomble05,Kallay05,Wloch05,Wloch06a,Kallay08,Eriksen14a,Eriksen15},
improve the computational efficiency 
compared to the full CCSD, CCSDT, and CCSDTQ models
and are useful practical approaches. 
For example, 
the CCSD augmented with a non-iterative treatment of triple excitations [CCSD(T)] \cite{Raghavachari89,Bartlett90,Stanton97} method
and the CCSDT augmented with a non-iterative treatment of quadruple excitation [CCSDT(Q)] \cite{Bomble05} method
feature balanced computational cost and accuracy and thus are being widely used in molecular calculations. 
Analytic-derivative formulations for CC methods \cite{Adamowicz84,Scheiner87,Salter89a,Salter89b,Koch90b,Gauss91a,Gauss91b,Gauss91c,Rendell91,CCptgrad,CCptgradLee,Watts92,Watts93,Gauss95a,Gauss95b,Gauss97a,Stanton00,Rauhut01,Gauss00,Gauss02a,Gauss02b,Kallay03,Kallay04a,Bozkaya16,Gyorffy18,Feng19,Matthews20c,Schnack-Petersen22} have been developed to enable
efficient CC calculations of molecular properties.
The availability of analytic CC energy derivatives greatly facilitates chemical and spectroscopic applications of CC methods.

 Relativistic effects \cite{Pyykko88,Dyall07,Autschbach12,Reiher14} play an important role in heavy-atom-containing molecules, including determining the level structure and measurable consequences of new fundamental particles and forces. For example, 
 the enhancement factors for the electron EDM in atoms \cite{Flambaum1976} and molecules \cite{Sushkov1978} 
 are intrinsically relativistic. Their values may exceed 1,000 when calculated with accurate treatments of relativistic effects, but they vanish in the non-relativistic limit.
 Relativistic effects also enhance the NSM sensitivity factors in heavy atoms by an order of magnitude. \cite{Sushkov84}
 Relativistic CC methods, \cite{Liu21a} i.e., CC methods combined with relativistic Hamiltonians,
 can provide accurate treatments of relativistic and electron-correlation effects 
 for single-reference molecular states containing heavy atoms; 
 they are useful tools for studying heavy-element chemistry and spectroscopy.
 Scalar-relativistic CC calculations \cite{Kaldor94,Leininger96,Fleig05,Cheng11,Cheng11b,Kirsch19} 
 retain the spin symmetry and are essentially as efficient
 as the corresponding non-relativistic CC calculations. 
 On the other hand, relativistic spin-orbit CC (SO-CC) methods \cite{Visscher95,Visscher96,Lee98,Eliav98,Wang08,Wang08b,Nataraj10,Liu18b} with variational treatments of spin-orbit coupling, 
 including spinor-based relativistic CC methods,
 are computationally more expensive 
 because
 of spin-symmetry breaking.
 While these methods were most often used in calculations of atoms and small molecules,
 recent algorithmic and implementational advances 
extend the applicability of SO-CC methods to larger molecules. \cite{Liu18b,Liu21a,Pototschnig21,Zhang24}
 
 \subsection{Atoms and Molecules as Probes for New Physics}
 
 Precision searches for energy-level splittings due to time-reversal symmetry-violating interactions
provide a promising approach to search for undiscovered fundamental physics beyond the Standard Model (BSM). \cite{DeMille15,DeMille17,Safronova18,Cairncross19}
In particular, heavy-atom-containing polar molecules possess orders of magnitude larger time-reversal symmetry-violating sensitivity parameters compared to atoms due to their ability to be polarized in lab fields \cite{Safronova18}; such molecules can serve as sensitive probes of BSM physics. \cite{Sandars65,Sandars67} As a prominent example, the use of molecules in searches for the electron's electric dipole moment (eEDM) has improved the upper bound for the eEDM value by several orders of magnitude \cite{Hudson02,Hudson11,Baron14,Andreev18,Cairncross17,Roussy23} compared to the earlier record obtained from atoms. \cite{Regan02} The ongoing work on improving existing experiments aims at orders of magnitude further improvement.\cite{Wu20,Ng22,Zhang22a,Hiramoto23,Popa24}
The EDM experiments may also be used to search for 
$P-T$ violating interactions between electrons and nucleons
in atoms and molecules mediated by new fundamental particles such as the axion,
which is one of the leading candidates for dark matter and was originally proposed
to solve the strong $CP$ problem. 
\cite{Stadnik15,Stadnik17,Stadnik18,Flambaum20,Flambaum20a,Flambaum20b,Maison21,Roussy21,Dalton2023enhanced}

The targeted interactions in the precision spectroscopy searches of new physics 
are related to intrinsic electronic-structure properties of atoms and molecules,
which can only be obtained from first-principle calculations until symmetry violations are observed.
For example, the searches for the eEDM in atoms and molecules probe an energy shift that scales with both the eEDM--an intrinsic property of the electron--and a sensitivity parameter known as the ``effective electric field'' \cite{Safronova18,Ginges04,Flambaum1976,Sushkov1978,Johnson86}
that depends on the atomic or molecular state used in the search, often referred to as the ``science state.''
First-principle calculations for sensitivity parameters of the science states 
play an essential role in selecting candidate molecules 
and in interpreting measurement results.

Science states of atoms and molecules in the precision measurement searches for BSM physics 
often comprise single-reference electronic states, i.e., electronic states dominated by
a single electron configuration, 
since their relatively simple structure makes them attractive candidates for experiments. 
 For example, among the science states being used in eEDM searches, 
 the $X^2\Sigma$ state of YbF is dominated by a Yb$^{+}$[$6s^1$]F$^{-}$ configuration, 
 the $^3\Delta_1$ states of ThO and ThF$^+$ are dominated by a Th$^{2+}$[$7s^16d^1$]O$^{2-}$ configuration and a Th$^{2+}$[$7s^16d^1$]F$^{-}$ configuration, respectively,
 and the $^3\Delta_1$ state of HfF$^+$ is dominated by a Hf$^{2+}$[$6s^15d^1$]F$^{-}$ configuration.
Spinor-based relativistic coupled-cluster methods have served as important tools to provide accurate computed properties,
e.g., the effective electric field,for these states. \cite{Cossel12,Skripnikov13,Petrov14,Fleig14,Denis15,Skripnikov15a,Skripnikov16,Skripnikov17,Fleig17a,Fleig21,Zhang22a,Ng22}

The campaign to observe BSM physics in general calls for a broad search for time-reversal symmetry-violating interactions including effects in both the leptonic and hadronic sectors.~\cite{Safronova18}  In addition to the eEDM, atomic and molecular systems with a nuclear spin $I>0$ can exhibit $T$-violating energy shifts arising from the NSM, and systems with $I>1/2$ can exhibit $T$-violating shifts arising from the nuclear MQM (NMQM). Each nuclide of the appropriate spin possesses its own characteristic NSM and NMQM values, which can arise from a combination of many underlying $T$-violating parameters including quark EDMs, quark chromo-EDMs, pion-nucleon couplings, and the Standard Model $\theta_{\text{QCD}}$ parameters \cite{flambaum2020electric,flambaum2014time}.
For spherical nuclei, NSMs involve contributions from both valence and internal nucleons~\cite{Flambaum1986onthe}, whereas NMQM values are dominated by the valence nucleon~\cite{flambaum2014time}. Mechanisms exist to enhance both parameters in deformed nuclei. Quadrupole deformed nuclei exhibit NMQMs enhanced by 1--2 orders of magnitude because many nucleons can occupy open shells~\cite{flambaum1994spin,flambaum2014time,lackenby18}. Octupole deformation leads to additional mechanisms that increase the NMQM by $\sim$1 order of magnitude compared to spherical nuclei~\cite{flambaum2022enhanced} and the NSM by $\sim$1--3 orders of magnitude~\cite{Sushkov84,Spevak1997enhanced,haxton1983enhanced,flambaum2020electric} due to symmetry-violating-interactions that mix the ground and nearly degenerate excited nuclear states. 
An axion dark matter field may also produce oscillating NSMs and NMQMs. \cite{Stadnik14,Abel17,Aybas21} 
There is growing interest in measurements of both NSMs and NMQMs~\cite{Fan21,Klos22,yu2021probing,Maison20,Grasdijk21,Denis20,udrescu2024precision,Arrowsmithkron23}. In this vein, an important application of spinor-based relativistic CC methods is the
 prediction of time-reversal symmetry-violating sensitivity parameters
 in atoms and molecules for probes of new physics beyond the Standard Model (BSM). 
\cite{Sahoo06,Skripnikov13,Kudashov13,Fleig13,Fleig14,Kudashov14,Abe14,Sasmal15c,Prasannaa15,Denis15,Skripnikov15,Skripnikov15b,Skripnikov16,Sasmal16,Fleig16,Fleig17,Fleig17a,Skripnikov17,Das17,Abe18,Maison19,Denis19,Prasannaa19,Sunaga19,Fazil19,Abe20,Skripnikov20,Maison20,Denis20,Talukdar20,Talukdar20a,Fleig21,Haase21,Zhang21a,Zakharova21,Hubert22,Maison22,Chamorro22,Oleynichenko22,Marc23,Gaul24}

\subsection{Molecular Structure Calculations Relevant to NSMs}

In this work, using analytic relativistic coupled-cluster gradient techniques, we perform molecular structure calculations relevant to experimental searches for NSMs in heavy nuclei, including the exceptionally important case of octupole deformed nuclei. Polar molecules are sensitive probes of the NSMs in their constituent nuclei.~\cite{Dzuba02} As described in detail in the theory section, 
an NSM produces an energy shift in an atom or molecule that scales with the electron density gradient near the nucleus. A nonzero density gradient at the nucleus is primarily present in $s-p$ hybridized wave functions.
While an electron density gradient can be achieved in atoms by polarizing electronic orbitals in an external electric field, in practice only modest density gradients at a nucleus can be experimentally achieved due to the infeasibly large applied electric fields that are required for full polarization. On the other hand, electronic orbitals in polar molecules naturally exhibit large density gradients at heavy atomic nuclei, and the molecules must only be oriented in the lab frame in order to provide full access to the NSM interaction. The ratio of the $T$-violating energy shift of a fully oriented molecule and the corresponding NSM is referred to as the ``molecular sensitivity factor.'' Note that the overall sensitivity of a molecular species to BSM physics relies on understanding both the molecular sensitivity factor, which is discussed in this work, and the nuclear sensitivity factor, which is not.\cite{Alarcon22}

Such a molecular sensitivity factor for the NSM interaction in a polar molecule is an intrinsic electronic-structure property and at the present stage can only be obtained from electronic-structure calculations. 
In the relativistic CC \cite{Kudashov13,Skripnikov20,Abe20,Gaul24} and MRCI \cite{Hubert22,Marc23} calculations reported previously, these parameters have been calculated as first derivatives of electronic energies using numerical differentiation  of electronic energies \cite{Kudashov13,Skripnikov20,Gaul24} or approximate analytic-gradient formulations. \cite{Abe20,Hubert22,Marc23} These calculations are challenging because of the high computational cost of spinor-based relativistic CC and MRCI methods. In addition, since the NSM sensitivity factor probes the derivative of the electron density distribution in the vicinity of a heavy atom, tedious numerical differentiation procedures are required to ensure numerical stability in the finite-difference procedure. 

The evaluation of NSM sensitivity factors as
analytic relativistic coupled-cluster energy gradients
is a promising scheme to expedite the calculations.
An analytic coupled-cluster gradient calculation
is around 2-3 times as expensive as a corresponding energy calculation \cite{Stanton00}
and can provide all first-order properties including various symmetry-violation
sensitivity parameters. 
It is also free from the complication of the numerical-differentiation procedure
and hence is of a ``black-box'' nature.
We have recently developed analytic gradients for exact two-component (X2C) CCSD, CCSD(T),
and equation-of-motion CCSD (EOM-CCSD) methods. \cite{Liu21,Zheng22,Zhang23a}
We have demonstrated calculations of effective electric fields based on analytic X2C-CCSD and CCSD(T) gradient techniques. \cite{Zhang21a} 
In this work, we extend the applicability
to calculations of NSM molecular sensitivity factors,
aiming to transform these previously complicated calculations 
into routine applications of a black-box nature.

\section{Methods} \label{sec:theory}

The contribution from the nuclear Schiff moment (NSM) interaction 
to the molecular energy is given by \cite{Dzuba02}
\begin{eqnarray}
E_{\text{NSM}}= \frac{3ke}{B} \langle \Psi |S_z \sum_{j} (z_j-z_N) \rho_N(\vec{r}_j;\vec{r}_N)| \Psi \rangle,
\label{eq:ENSM}\end{eqnarray}
in which $|\Psi\rangle$ is the molecular wave function, 
$S_z$ is the component of the NSM 
along the molecular axis,
$\rho_N(\vec{r};\vec{r}_N)$ is the nuclear charge density at $\vec{r}$ of the active nucleus centered
at $\vec{r}_N$,
and $\vec{r}_j$ represents the position of electron $j$. The constant $B$ is a property of the nuclear distribution, described below. In the following we adopte the natural units of $k=e=1$, where $k$ is Coulomb's constant and $e$ is the elementary charge. The expression above can be interpreted as the electrostatic interaction between the electrons and an electric field approximately along the $\mathrm{sgn}(S_z)\hat{z}$ direction within the nucleus. \cite{Flambaum02}

In our calculations we adopt 
a finite nuclear model using 
a single Gaussian function to represent the nuclear charge distribution. 
Within this model, a finite nuclear charge distribution for nuclear charge $Z$ 
with a radial distribution
of a Gaussian function
\begin{eqnarray}
\rho_N(r)=\rho_0 e^{-\xi{r^2}}~,~\rho_0=Z\left(\frac{\xi}{\pi}\right)^{3/2}
\end{eqnarray}
centered at $\vec{r}_N$ 
can be written as
\begin{eqnarray}
\rho_N(\vec{r};\vec{r}_N)=\rho_0 e^{-\xi(\vec{r}-\vec{r}_N)^2}. 
\label{eq:nuclear-density}\end{eqnarray} 
$B$ is defined as
\begin{eqnarray}
B=\int_0^\infty \rho_N(r) r^4 d{r},
\end{eqnarray}
and takes the simple form \cite{Marc23}
\begin{eqnarray} 
B=\frac{3}{8\pi\xi}
\end{eqnarray}
when this Gaussian nuclear distribution is used.

The effective Hamiltonian for the NSM interaction can thus be written as
\begin{eqnarray}
\hat{H}^{\text{eff}}_{\text{NSM}}=S_z W_{\text{NSM}},
\end{eqnarray}
with the NSM sensitivity factor $W_{\text{NSM}}$ defined as
\begin{eqnarray}
W_{\text{NSM}}&=&\frac{3}{B} \langle \Psi |\sum_{j} (z_j-z_N) \rho_N(\vec{r}_j;\vec{r}_N)| \Psi \rangle. \label{NSMEXP}
\end{eqnarray}
Equivalently, for the nuclear density described by Eq.~\ref{eq:nuclear-density}, $W_{\text{NSM}}$ can be written in terms of the first derivative of the ``effective density'' with respect to the coordinate of the center of the nuclear charge distribution 
\begin{eqnarray}
W_{\text{NSM}}= 4\pi \left.\frac{\partial \bar{\rho}_e[\vec{r}_C]}{\partial z_C}\right\vert_{\vec{r}_C=\vec{r}_N}, \label{NSMDEN}
\end{eqnarray} 
in which the effective density, $\bar{\rho}_e[\vec{r}_C]$, at a position $\vec{r}_C$ 
is defined in the same way as 
in the calculations of isomer shifts in M{\"o}ssbauer spectroscopy,
\cite{Filatov07,Knecht11}
\begin{eqnarray}
\bar{\rho}_e[\vec{r}_{C}]=\langle \Psi | \sum_{j} \rho_N(\vec{r}_j;\vec{r}_{C}) | \Psi \rangle.
\end{eqnarray}
In the present calculations the origin of the coordinate system is chosen as the center of mass of a molecule.
The coordinates of the active nucleus always assume positive values, namely, $z_N>0$.
Therefore, a negative (positive) value for $W_{\text{NSM}}$ represents excessive (deficient) effective electron density in the inter-atomic bonding region. 

As shown in Eq. (\ref{NSMEXP}), $W_{\text{NSM}}$ is an expectation value of the molecular wave function.
It can be evaluated as the first derivative of the electronic energy 
by augmenting the molecular Hamiltonian $\hat{H}_0$ with the NSM interaction $S_z \hat{H}_{\text{NSM}}$
\begin{eqnarray}
\hat{H}=\hat{H}_0+S_z \hat{H}_{\text{NSM}},~ \hat{H}_{\text{NSM}}=\frac{3}{B} \sum_{j} (z_j-z_N) \rho_N(\vec{r}_j;\vec{r}_N),
\end{eqnarray}
and evaluating the first derivative of the electronic energy with respect to
$S_z$, namely,
\begin{eqnarray}
W_{\text{NSM}}=\left.\frac{\partial E}{\partial S_z}\right\vert_{S_z=0}.
\end{eqnarray}
Within the analytic relativistic coupled-cluster gradient formulation,
$W_{\text{NSM}}$ is evaluated in a straightforward manner 
by contracting the relaxed coupled-cluster one-electron density matrix
with one-electron integrals for $\hat{H}_{\text{NSM}}$.
This avoids the complication in numerical differentiation
and enables black-box calculations of this parameter.

The calculation of $W_{\text{NSM}}$ has been implemented in the CFOUR program package
\cite{Matthews20a,CFOURfull}
using the analytic-gradient implementation for exact-two-component (X2C) CCSD and CCSD(T) methods.
\cite{Liu21}
We have used the analytic X2C-CC gradient module developed earlier to 
calculate the relaxed X2C-CCSD and CCSD(T) one-electron density matrices 
and to construct X2C derivative integrals \cite{Cheng11b} from four-component property integrals. 
The new implementation work is focused on the evaluation of the 
four-component integrals for the operator $\hat{H}_{\text{NSM}}$.
To ensure numerical stability, quadruple precision has been adopted in the evaluation
of these integrals. 
The correctness of this implementation has been verified in two ways.
First, since $\sum_{j} \rho_N(\vec{r}_j;\vec{r}_N)$ is involved in the calculations
of effective densities in M{\"{o}}ssbauer spectroscopy, 
we have verified the correctness for the evaluation of four-component integrals for $\sum_{j} \rho_N(\vec{r}_j;\vec{r}_N)$
by comparing the computed effective densities to reference values in the literature. \cite{Knecht11}
Then the multiplication with $z_j-z_N$ is straightforward to add in the program module to obtain integrals for $W_{\text{NSM}}$.
Second, we perform numerical differentiation of effective densities to obtain $W_{\text{NSM}}$ values
based on Eq. (\ref{NSMDEN}) and compare the results to $W_{\text{NSM}}$ values obtained as analytic energy derivatives.
In the finite-difference procedure, we calculate effective densities on nine grid points [$\vec{r}_C-\vec{r}_N$=(0, 0, $n\times 10^{-8}$), n=-4, -3, $\cdots$, 3, 4] relative to the active nucleus.
We fit the results to a fourth-order polynomial to extract the first derivative of the effective density with respect to
the coordinate and then convert it into $W_{\text{NSM}}$ using Eq. (\ref{NSMDEN}). 
The $W_{\text{NSM}}$ values thus obtained via numerical differentiation agree well
with the values obtained from the analytic evaluation, with discrepancies below 0.02\%.  %

All of the calculations presented here have used the X2C Hamiltonian \cite{Dyall01,Ilias07,Liu09} with atomic mean-field integrals \cite{Hess96a} (the X2CAMF scheme) \cite{Liu18,Zhang22}
to treat relativistic effects. Gaussian nuclear distributions as parametrized in Ref. \cite{Visscher97} have been used throughout the calculations. 
We have adopted the X2CAMF scheme based on the Dirac-Coulomb-Breit Hamiltonian \cite{Zhang22} in the calculations, except that we have performed one calculation using the X2CAMF scheme based on the Dirac-Coulomb Hamiltonian [the X2CAMF(DC) scheme] \cite{Liu18} for comparison to a four-component DC calculation.
The spinor-based X2CAMF-CC calculations have been expedited using the recent implementation of atomic-orbital based algorithms. \cite{Liu18b}
The CC calculations have correlated 
the valence and semicore electrons
defined in the supporting information. 

Since the NSM sensitivity factor probes the electron-density gradient in the vicinity of a heavy nucleus,
an accurate calculation requires the basis sets to describe the wave functions close to the nucleus with high accuracy.
Standard basis sets optimized for chemical properties are insufficient for this purpose; 
even the uncontracted versions of standard basis sets have been shown to exhibit instability
in calculations. \cite{Hubert22} 
The present study has employed even-tempered series of $s$- and $p$-type functions
for the targeted atoms
to ensure a unified description for core and valence regions. 
A general scheme to construct basis sets is to take the uncontracted version of a standard basis set and replace
the portion of $s$- and $p$-type functions exhibiting exponent intervals larger than 2.5
with even-tempered basis (ETB) sets with
exponents given by $\{\alpha \beta^{m-1}, m=1,\cdots, n\}$, in which 
$\beta=2.5$, and ensure that 
the largest exponent in the ETB series is greater than $10^8$ to obtain a good description of the core region. 
We have used basis sets of triple-zeta quality in this procedure to obtain the ``ETB0'' sets.
For example, the ETB0 set for thorium is obtained by taking the uncontracted ANO-RCC basis set \cite{Faegri01,Roos05a}
and replacing the $s$-type functions having exponents larger than 78732.668 with 
an even-tempered series with exponents $\{\alpha \beta^{m-1}, \alpha=78732.668, \beta=2.5, m=2,\cdots, 9\}$
and $p$-type functions having exponents larger than 4672.81862 with
an even-tempered series with exponents $\{\alpha \beta^{m-1}, \alpha=4672.81862, \beta=2.5, m=2,\cdots, 12\}$. 
The ETB0 sets have been used in most of the calculations presented here.
In order to study the remaining basis-set effects beyond the ETB0 sets,
we have carried out calculations using basis sets with additional tight functions and high angular momentum functions. 
The basis sets and molecular structures 
used in the present calculations are documented in the Supporting Information.

\section{Results and discussions}

\subsection{Benchmark studies: Basis-set convergence, electron-correlation contributions, and comparison with literature}

Let us first study the dependence of the computed 
NSM sensitivity factors (the $W_{\text{NSM}}$ values)
to the choice of basis sets.
We start with a study of the contributions from additional tight $s$- and $p$-type functions.
We add three additional tight $s$- and $p$-type functions to the ETB0 sets to obtain the ETB0+SP3 sets.
As shown in Table \ref{tab1}, the inclusion of three additional tight $s$- and $p$-type functions 
makes small contributions to the computed $W_{\text{NSM}}$ values.
In the case of FrLi, the difference between ETB0+3SP and ETB0 results amounts to 
-60 a.u. at the HF level, less than 0.03\% of the total value, and 2 a.u. in the electron-correlation contribution ($W_{\text{NSM}}$ is presented in atomic units of $E_{h}(ea_{0}^{3})^{-1}$ throughout). 
The corresponding contributions in RaF and ThO are also insignificant.
Therefore, the ETB0 sets have sufficient tight $s$- and $p$-type functions for accurate calculations of 
the NSM sensitivity factors. 
We then investigate the use of even-tempered $d$-type functions in the calculations.
We replace the $d$-type functions in the ETB0 sets with even-temper sets of $d$-type functions
to obtain the ETBSPD sets. 
The differences between ETBSPD and ETB0 results amount to
0.04\% of the total value in the case of FrLi and around 1\% for RaF and ThO. 
They appear to be insignificant compared with other sources of errors, 
e.g., the electron-correlation contributions; 
it is thus sufficient to 
use the $d$-type functions in the standard basis sets for the calculations of $W_{\text{NSM}}$. 
Finally, we include higher angular momentum functions to enhance the quality of the basis sets to quadruple-zeta quality
and obtain the ETBQZ sets. This also introduces relatively small corrections. 
The differences between ETBQZ and ETB0 results are smaller than 0.5\% of the total values. 
Therefore, these benchmark results show that the ETB0 sets have sufficient flexibility 
to provide reliable $W_{\text{NSM}}$ values. 

\begin{table}
	\caption{
Computed $W_{\text{NSM}}$ values (a.u.) using the X2CAMF scheme to treat relativistic effects.
Enclosed in the parentheses are the electron-correlation contributions at the CCSD level, namely,
the differences between CCSD and HF results. 
	 } 
\center{
\begin{tabular}{ccccccc}
 \hline \hline
   &   \multicolumn{2}{c}{FrLi}  &   \multicolumn{2}{c}{RaF}     &  \multicolumn{2}{c}{ThO (H$^3\Delta_1$)}  \\        
      \hline
   & HF & CCSD & HF & CCSD & HF & CCSD \\
   \hline
  ETB0          &  -25006 & -23063   &   -23147 &  -20254    & -32243 & -25772 \\
                     &              & (1943)    &               &   (2893)    &             &   (6471) \\
  ETB0+3SP &  -25068 & -23123   &    -23153 & -20258     & -32356 & -25881  \\
                      &              & (1945)    &               &   (2895)    &             &   (6475) \\
  ETBSPD    &  -24968 &  -23027   &   -22914 &  -20027     & -31904 & -25465 \\
                      &              & (1941)    &               &   (2887)    &             &   (6439) \\
  ETBQZ       &  -24936 & -23079   &   -23247 & -20354     & -32368 & -25846  \\ 
                      &              & (1857)    &               &   (2893)    &             &   (6422) \\      
 \hline\hline                                            
 \end{tabular}}
 \label{tab1}
\end{table}

As expected, electron correlation makes important contributions to 
the $W_{\text{NSM}}$ values. 
As shown in Table \ref{tab1}, the electron-correlation contributions at the CCSD level
amount to around 8\%, 14\%, and 25\% of the total $W_{\text{NSM}}$ values
in FrLi, RaF, and ThO, respectively. 
It is thus necessary to include electron-correlation contributions to obtain accurate $W_{\text{NSM}}$ values.
Relativistic coupled-cluster methods provide robust treatments of electron correlation
and are methods of choice for reliable prediction of $W_{\text{NSM}}$ values.
Meanwhile, we note that the HF calculations can provide qualitatively correct results 
that serve as useful initial estimates for the magnitudes of the $W_{\text{NSM}}$ values.
Based on the present CC results, we find that
relativistic density-functional theory calculations \cite{Gaul24} provide reasonably good $W_{\text{NSM}}$ values. 

\begin{table}
	\caption{
Computed $W_{\text{NSM}}$ values (a.u.) for the closed-shell
electronic ground states of heavy-atom-containing molecules
using the X2CAMF scheme to treat relativistic effects.
The ETB0 basis sets were used for the active atoms (Ac, Eu, Tl, Fr, Ra).
The uncontracted aug-cc-pVTZ basis sets were used for the light atoms (F, N, O, S, H),
and the uncontracted ANO-RCC basis set was used for Ag.
{{
Note that a scaling factor of 1/1.13 \cite{Flambaum20d,Skripnikov20} should be applied to the $W_{\text{NSM}}$ value for TlF in Ref. \cite{Abe20} obtained using point 
nuclear model when comparing it with the other results
obtained using finite nuclear charge distribution.}} 
	 } 
\center{
\begin{tabular}{ccccccc}
 \hline \hline
   & HF & CCSD & CCSD(T) & Literature \\
   \hline
    AcF         &  -9787      &   -7573  & -7399    & -8240 \cite{Skripnikov20} \\   
    AcN         &  -44577    &   -38874  & -37136    & -46295 \cite{Skripnikov20} \\   
    AcO$^+$         &  -57778    &   -48857  & -46936    & -58461 \cite{Skripnikov20} \\    
    ThO         &   -4351     &  -1684 &  -2330  &-17085 \cite{Skripnikov20} \\
    EuO$^+$         &  -10775   &   -9199  & -9145    & -11677 \cite{Skripnikov20} \\ 
    EuN         &  -7300   &   -7086  & -7280    & -10419 \cite{Skripnikov20} \\ 
    TlF         &  45021   &   35319  & 33279   & 37192\cite{Skripnikov20}/41136 \cite{Abe20}/39967 \cite{Hubert22} \\    
    FrAg       &  -30815   &   -29370  & -28737   & -30168 \cite{Marc23} \\  
    RaO       &  -57413   &   -46878 & -43994   & -45192 \cite{Kudashov13} \\       
    RaSH$^+$        &  { {-48590}}     & { {-44275}} & { {-43360}}   &-45060 \cite{Gaul24} \\           
 \hline\hline                                            
 \end{tabular}}
 \label{tab2}
\end{table}

Table \ref{tab2} provides a comparison of $W_{\text{NSM}}$ values computed in the present study to
available calculations in the literature. 
The present results for AcF, AcN, AcO$^+$, EuO$^+$, EuN, and TlF
are around 20\% smaller than the values reported in Ref. ~\onlinecite{Skripnikov20}.
The only major discrepancy that we observe is that 
the present results for the $^1\Sigma^+$ state of ThO are significantly smaller than
those reported in Ref. ~\onlinecite{Skripnikov20}.
For example, the present X2CAMF-HF value of -4351 a.u. is an order of magnitude smaller than
the HF value of -20333 a.u. in Ref. ~\onlinecite{Skripnikov20}.
We performed a HF calculation using the X2CAMF(DC) scheme and obtained a value of -4621 a.u.. 
This X2CAMF(DC)-HF value agrees well with 
a value of -4814 a.u. obtained from four-component DC-HF calculations by Skripnikov
using the same molecular structure and basis sets. 
The discrepancy between these results obtained from all-electron relativistic two- and four-component calculations
and the results in Ref. ~\onlinecite{Skripnikov20} obtained using a two-step scheme
is thus attributed to 
the approximations in the two-step scheme. 

{{The X2CAMF-HF value of 45021 a.u. for TlF agree closely with the four-component value of 45419 a.u. in Ref. \cite{Hubert22},
which also has a thorough investigation on basis-set effects. 
The HF values in Refs. ~\onlinecite{Skripnikov20} and ~\onlinecite{Abe20} are around 7\% larger.
The X2CAMF-CC electron correlation contribution of -11741 a.u. are consistent with
the corresponding relativistic CC results in Resf. ~\onlinecite{Skripnikov20} and ~\onlinecite{Abe20}.
For comparison, the MRCI electron-correlation contribution of -5452 a.u. \cite{Hubert22} is substantially smaller in terms
of the absolute magnitude. }}

The present results for RaSH$^+$ and RaO agree well 
with available relativistic coupled-cluster calculations, \cite{Kudashov13, Gaul24} with
discrepancies below 4\% of total values. 
The present results for FrAg also agree well
with recent relativistic HF and MRCI calculations. \cite{Marc23}
Here the HF value of -30815 a.u. in the present calculation differs from
the HF value of -31350 a.u. reported in Ref. \cite{Marc23} 
by around 500 a.u., less than 2\% of the total value. 
The difference between the electron-correlation contributions 
amounts to around 800 a.u., which can be attributed
to the difference of CC and MRCI in treating electron correlation.


\subsection{Chemical Mechanisms Contributing to $W_{\text{NSM}}$}

A $W_{\text{NSM}}$ value is proportional to the gradient of the effective electron density 
at the position of the active nucleus.
It has been pointed out in Ref. ~\onlinecite{Skripnikov20} that the electron-density gradient at the position of a heavy atom may have two competing contributions. 
The electronegative ligand draws electron density from the heavy atom toward the ligand.
 Meanwhile, the net negative charge in the ligand tends to push the electron density 
 at the heavy atom away from the ligand.
 Here we specify these two competing types of contributions based on simple chemical concepts.
 The first one is readily attributed to polar covalent chemical bonds between 
 the heavy metal and an electronegative neighboring atom or functional group.
In all of the calculations presented here, 
the coordinate of the active nucleus along the molecular axis assumes positive values.
The polar covalent chemical bonds draw electron density toward the inter-atomic bonding region and thus makes a contribution with a negative sign to $W_{\text{NSM}}$. 
The second contribution can be attributed to the back-polarization 
of non-bonding orbitals of the heavy atom, especially
the non-bonding valence $s$-type orbitals, due to
the ligand field. 
This back-polarization of $s$-type orbitals creates an electron-density deficiency
in the inter-atomic bonding region and makes a contribution with a positive sign to $W_{\text{NSM}}$. 

We plot the relevant polarized molecular orbitals to illustrate
the first mechanism. As shown in Figure 1, a polar molecular orbital in TlF represents a typical $\sigma$ bonding orbital involving
Tl $6s$ (28\%), Tl $6p$ (2\%), Tl $5d$ (3\%), and F $2p$ (67\%) contributions. 
A polar molecular orbital in AcF that makes a significant negative contribution to the NSM value
involves Ac $7p$ (12\%), Ac $6d$ (6\%) and F $2p$ (78\%) contributions. 
These orbitals draw electron densities toward the
more electronegative fluorine atom. 
The present illustration is based on Hartree-Fock orbitals. 
It might be of interest for future work to perform bond analysis for these molecules
and study the relevance of covalency in the bonding
to the NSM sensitivity parameters.
The second mechanism may not be as well known as the first one,
while it can be illustrated well using the back-polarized non-bonding orbitals.
The back-polarized Ac $7s$ orbitals in AcF and Tl $6s$ orbitals in TlF are plotted in Figure 1. 
They lead to an electron density deficiency in the inter-atomic region
and contribute significant positive values to the NSM sensitivity parameters.
We mention that molecular electric dipole moment values also have these two competing types of contributions.
This has been demonstrated, for example, in joint experimental-computational work on thorium halides \cite{Nguyen19}
and in the computational study in Ref. ~\onlinecite{Skripnikov20}. \\

\scalebox{0.4}{ \put(-10.0,0.0){\includegraphics{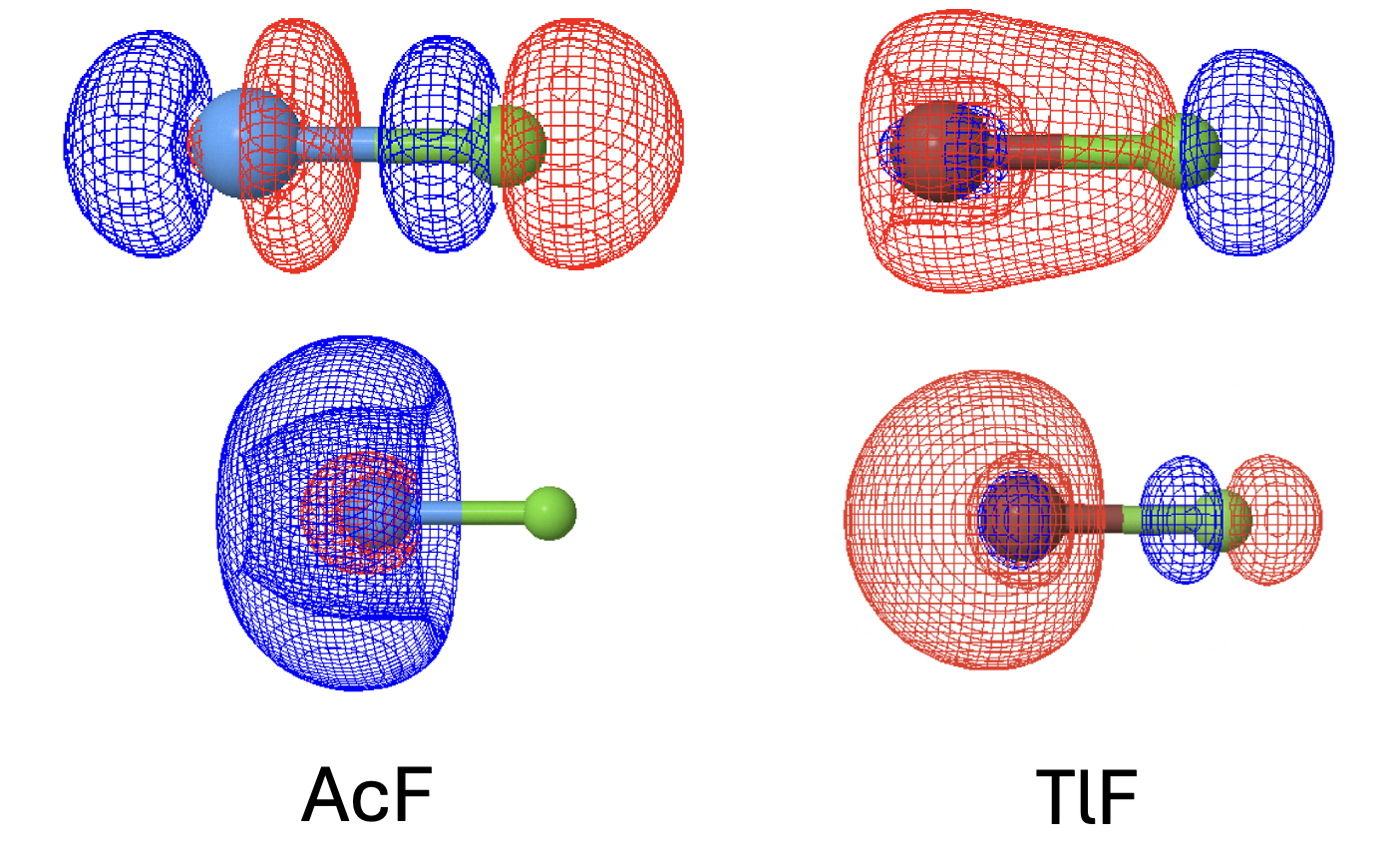}} }

Figure 1: The polar orbitals in AcF and TlF (top) that make significant negative contributions to NSM sensitivity parameters and back-polarized Ac $7s$ orbital in AcF and Tl $6s$ orbital in TlF (bottom) that make positive contributions. Isosurfaces with an isovalue of $0.04$ for the absolute value of the wave function are plotted. The red and blue colors represent opposite signs of the wave function values. \\

The relative magnitudes of the $W_{\text{NSM}}$ values in AcF, AcN, and AcO$^+$ can be explained using
these two mechanisms. The relatively small absolute magnitude of $W_{\text{NSM}}$ in AcF comes from
a major cancellation between these two contributions. 
Each electron in the bonding orbital shown in Figure 1 contributes a value of -22095 a.u.
to the NSM value. 
The two non-bonding Ac $7s$ electrons are significantly back-polarized. 
Each of them contributes a value of 14085 a.u. to $W_{\text{NSM}}$ at the HF level.
In contrast, AcN and AcO$^+$ have no non-bonding Ac $7s$ electrons.
The contributions from back-polarized Ac $7s$ electrons in AcF thus are responsible for most of 
the differences between the $W_{\text{NSM}}$ value in AcF and those in AcN and AcO$^+$. 

Interestingly, TlF is an example for particularly significant back-polarization of valence $s$-type orbitals, 
as shown in Figure 1. Each back-polarized $6s$ electron of Tl contributes
a value of as large as 32655 a.u. to $W_{\text{NSM}}$. 
Furthermore, the difference in electronegativity between Tl and F is not as
large as that between Ac and F in AcF; the contributions from the first mechanism due to polar covalent chemical bonds is relatively small in TlF.
The bonding orbital shown in Figure 1 contributes only -12583 a.u. to the NSM value. 
These together lead to a $W_{\text{NSM}}$ value of a peculiar positive sign in TlF. 

In summary, from the perspective of rational design, an ideal NSM-sensitive species would maximize the gradient of electron density at the heavy (ideally octupole-deformed) nucleus, which is achievable in a number of ways. One option is to select molecules with an electronegative ligand that withdraws electron density from the heavy nucleus, maximizing the (negative) contribution due to the polar covalent bond. This can be achieved in molecules with large electronegativity differences between the bonded atoms, and in cases where the heavy atom has empty non-bonding orbitals. Another option would be to select molecules with significant back-polarization of orbitals centered on the heavy atom, increasing the (positive) contribution to $W_\text{NSM}$. This can be achieved by selecting bonding partners with similar electronegativities, and doubly occupied non-bonding orbitals that undergo significant back-polarization. Competition between these two effects can lead to values of $W_\text{NSM}$ with reduced magnitude. In the following we present computed NSM sensitivity factors for thorium-, radium-, and dysprosium-containing molecules, together with analysis of the computational results based on these
 chemical concepts. 

\subsection{$W_{\text{NSM}}$ values in ThO and ThF$^+$}

ThO and ThF$^+$ are already important molecules in the  search for the eEDM.
The work on precision measurement of ThO by the ACME collaboration has improved the upper bound for the eEDM value by two orders of magnitude compared to earlier records. \cite{Baron14,Andreev18}
Ongoing and future measurements seek to further improve
the sensitivity to the eEDM. \cite{Wu20,Hiramoto23}
ThF$^+$ is the molecule of choice for the third-generation JILA eEDM measurement. \cite{Gresh16,Ng22}
Because the $^3\Delta_1$ state of ThF$^+$ has significantly longer coherence time and larger effective electric field than that of HfF$^+$, a precision measurement using ThF$^+$ has the potential to improve the current leading sensitivity to the eEDM set with HfF$^+$. \cite{Cairncross17,Roussy23}
These molecules are potential candidates for NSM searches, since
$^{227}$Th and $^{229}$Th have been proposed to have octupole deformation. \cite{Liang1995,Hammmond2002,ruchowska2006nuclear,gulda2002nuclear,chishti2020direct}
Therefore, it is of interest to investigate the possibility of using these molecules in the search for NSMs. 

\scalebox{0.5}{ \put(-10.0,0.0){\includegraphics{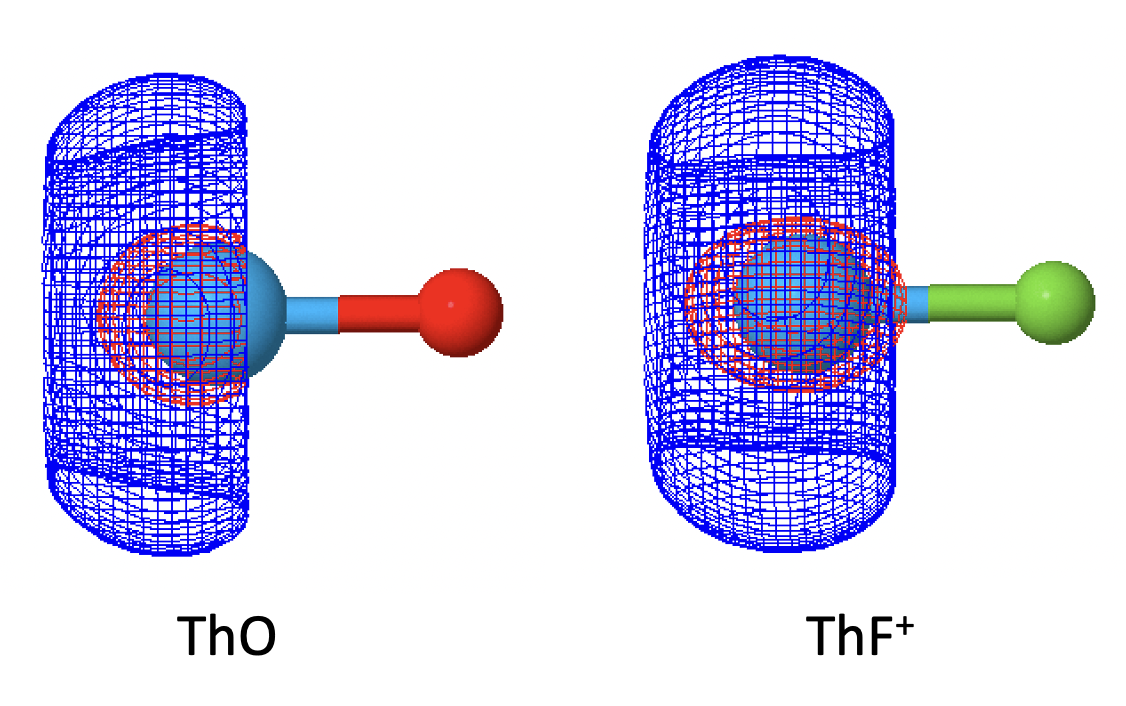}} }

Figure 2: Back-polarized Th $7s$ orbital in the $^1\Sigma^+$ states of ThO and ThF$^+$. Isosurfaces with an isovalue of $0.04$ for the absolute value of the wave function are plotted. The red and blue colors represent opposite signs of the wave function values.  \\
 
The lowest-lying electronic states in both ThO and ThF$^+$ are a closed-shell $^1\Sigma^+$ state (with Th electron configuration $7s^2$)
and an open-shell $^3\Delta_1$ state ($7s^1 6d^1$).
In ThO, the $^1\Sigma^+$ state is the ground state, and the $^3\Delta_1$ state is a metastable excited state.
By contrast, ThF$^+$ has a $^3\Delta_1$ ground state, while the $^1\Sigma^+$ state is located
approximately 500 cm$^{-1}$ higher in energy.
The computed $W_{\text{NSM}}$ values for both $^1\Sigma^+$ and $^3\Delta_1$ states of ThO and ThF$^+$ are
summarized in Table \ref{tab3}. 
ThO and ThF$^+$ have highly polarized chemical bonds drawing electron densities from Th toward O or F; 
all computed $W_{\text{NSM}}$ values take negative values.
Th $7s$ orbitals are back-polarized by F$^-$ or O$^{2-}$; Th $7s$ electrons contribute positive values to $W_{\text{NSM}}$
that partially cancel the contributions from the polarized chemical bonds.
Since a $^1\Sigma^+$ state has one more Th $7s$ electron than a
$^3\Delta_1$ state, the $W_{\text{NSM}}$ values for the  $^1\Sigma^+$ states
take smaller absolute values than those of the $^3\Delta_1$ states.
As shown in Figure 2, O$^{2-}$ back-polarizes the Th $7s$ orbital more effectively than F$^-$. 
The $W_{\text{NSM}}$ values in ThO thus show more pronounced cancellation and 
take smaller absolute values than the  
corresponding $W_{\text{NSM}}$ values in ThF$^+$. In particular,
the $^1\Sigma^+$ state of ThO has two heavily back-polarized Th $7s$ electrons and thus exhibits
a $W_{\text{NSM}}$ value significantly smaller than the other states. 

\begin{table}
	\caption{
Computed $W_{\text{NSM}}$ values using the X2CAMF scheme to treat relativistic effects.
The ETB0 basis sets were used for Th and Ra.
The uncontracted aug-cc-pVTZ basis sets for H, C, O, F were used 
in all calculations except that 
the calculations of RaCH$_3^+$, RaOCH$_3^+$, RaCH$_3$, and RaOCH$_3$ used the SFX2C-1e recontracted cc-pVTZ sets for O, C, and H.
The increments with respect to the preceding column are enclosed in the parentheses. } 
\center{
\begin{tabular}{ccccccc}
 \hline \hline
   & HF & CCSD & CCSD(T) \\ 
   \hline
      DyF (Dy$4f^{9}6s^2$)         &  -2578      &   -2113 (465) &  -2004 (108) \\  
      DyO  (Dy$4f^{9}6s^1$)         &  -9460      &   -7821 (1638) & -7431 (390)  \\        
      ThO    (X$^1\Sigma^+$)     &   -4351     &  -1684 (2667) &  -2330 (-646) \\
      ThO    (H$^3\Delta_1$)     &  -33176    &  -26576 (6600) &  -25384 (1192) \\
      ThF$^+$    (a$^1\Sigma^+$)     &  -25145      & -29811 (-4666)  &  -31460 (-1649) \\
      ThF$^+$    (X$^3\Delta_1$)     & -44988     &   -39927 (5061) &  -38403 (1524) \\
      RaF$^+$        &  { {-50534}}      &    { {-45056 (5478)}} &  { {-43911 (1145)}}   \\
      RaOH$^+$       &   { {-54950}}     &  { {-49189 (5761)}} &  { {-48103 (1086)}}  \\
      RaCH$_3^+$        &   { {-55741}}     &  { {-49758 (5983)}} &   { {-48517(1242)}}\\
      RaOCH$_3^+$        &   { {-55010}}     &  { {-49186 (5824)}}  &  { {-48053 (1133)}}  \\
      RaF       &  -23147      &   -20254 (2893) & -19674 (580)   \\
      RaOH        &  -26860     & -23397 (3463)  & -22877 (520) \\
      RaCH$_3$       &  -27908     &  -25202 (2706) &  -24695 (507) \\
      RaOCH$_3$  &  -26552     &   -23432 (3120)    &  -22928 (504)   \\
     \hline\hline                                            
 \end{tabular}}
 \label{tab3}
\end{table}

This back-polarization mechanism also explains the relative magnitude
of the effective electric fields in the $^{3}\Delta_{1}$ states of ThO and ThF$^+$.
The dependence of the effective electric field on $s-p$ mixing is directly analogous to the dependence of $W_{\text{NSM}}$ on the electron density gradient near the nucleus.
As mentioned above, the Th $7s$ orbital in ThO is back-polarized more significantly than
that in ThF$^+$ and has more contributions from the atomic Th $7p$ orbitals.
An inspection of the molecular spinor compositions shows that the back-polarized ``Th $7s$'' orbitals in ThO and ThF$^+$ have 7\% and 2.5\% ``Th $7p$'' contributions, respectively.
The back-polarized Th $7s$ electron in the $^3\Delta_1$ state of ThO
thus makes a larger contribution to the effective electric field than that in ThF$^+$, and the effective electric field in ThO is consequently larger than that in ThF$^+$. 
Note that while the effective electric field depends only on the open-shell electron(s), the NSM sensitivity factors have contributions from all electrons.
While the back-polarized Th $7s$ electron in ThO also makes a larger contribution in terms of absolute magnitude
for the NSM sensitivity factors,
it tends to cancel the contributions from polarized chemical bonds.
This results in a NSM sensitivity factor in ThO smaller in absolute magnitude
than in ThF$^+$.

\subsection{$W_{\text{NSM}}$ values for radium-containing molecules}

\scalebox{0.5}{ \put(-10.0,0.0){\includegraphics{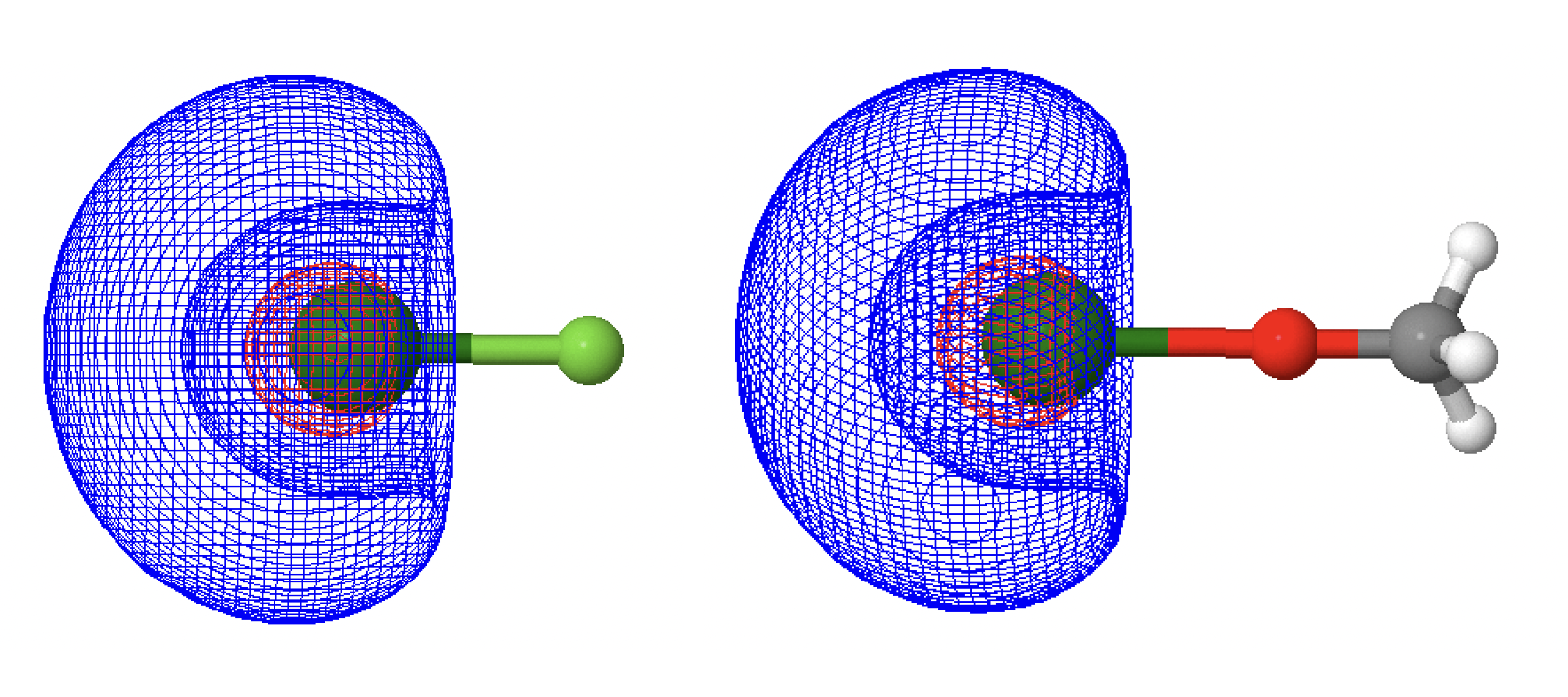}} }

Figure 3: Back-polarized Ra $7s$ orbitals in RaF and RaOCH$_3$. Isosurfaces with an isovalue of $0.04$ for the absolute value of the wave function are plotted. The red and blue colors represent opposite signs of the wave function values.  \\ 

Radium-containing polar molecules
have large internal electric fields and hence are expected 
to have favorable NSM sensitivity parameters.
Synthesis, cooling, and trapping of molecular species containing the octupole-deformed $^{225}$Ra nucleus \cite{Auerbach96}
thus have the potential to provide  
promising platforms for probes of the NSM interaction.
Significant progress has recently been reported in this direction.\cite{Arrowsmithkron23}
Sympathetic cooling of radium-containing molecular cations \cite{Fan21} 
and high-resolution laser spectroscopy for 
a radium-containing neutral molecule, RaF, \cite{GarciaRuiz20} have been reported. 
Spectroscopic measurements and relativistic electronic-structure calculations 
also support the laser-coolability of RaF and RaOH.  
\cite{Isaev10,Zaitsevskii22,Osika22,Athanasakiskaklamanakis23,Zhang23,Zhang23}
Relativistic electronic-structure calculations have also shown that radium-containing molecules exhibit favorable NSM sensitivity factors. \cite{Kudashov13, Gaul24}

Computational results of relativistic coupled-cluster calculations for a set of radium-containing molecules are summarized in Table \ref{tab3}.
All radium-containing molecular cations presented here are closed-shell species with
no non-bonding valence electrons located at the radium atom.
The $W_{\text{NSM}}$ values come entirely from the polar chemical bonds between radium and
(pseudo-)halogen ligands. As shown in Table \ref{tab3}, 
The $W_{\text{NSM}}$ parameters of these molecular cations take negative values with large absolute magnitude.
The neutral species presented here are all molecular radicals having one unpaired electron occupying a Ra $7s$ orbital,
which is significantly back-polarized by the ligand, as shown in Figure 3. 
While this property is favorable for laser cooling, the $W_{\text{NSM}}$ contribution from the unpaired electron in the back-polarized orbital tends to cancel the contribution from the polar chemical bonds.
Therefore, the $W_{\text{NSM}}$ parameters in these neutral molecules are smaller than those in the cations 
in terms of absolute magnitude, by roughly a factor of two. Nevertheless, all radium-containing molecular species studied here
exhibit significant molecular sensitivity factors for the NSM interaction; they are promising candidate molecules
for precision measurement probes of BSM physics in the hadronic sector and the strong CP problem. 

\subsection{$W_{\text{NSM}}$ values for Dy-containing molecules}

Notable nuclei with non-zero octupole deformation parameters, $\langle\beta_{3}\rangle\neq0$, are believed to exist only in radioactive atoms such as Rn--Pu~\cite{Dalton2023enhanced}. However, certain lanthanide nuclei also exhibit a dynamical, vibration-like octupole deformation such that $\langle\beta_{3}^{2}\rangle\neq0$. Since the enhanced NSM depends quadratically~\cite{Spevak1997enhanced} on the deformation parameter, $S\propto\beta_{3}^2$, these dynamically deformed species may also offer the possibility of highly sensitive measurements. For example, experiments using stable, dynamically octupole-deformed $^{153}$Eu$^{3+}$ ions embedded in a crystal are under development, with anticipated sensitivity to $CP$-violating physics several orders of magnitude beyond current bounds~\cite{Ramachandran2023nuclear,Sushkov23}. Octupole deformation in Dy has also been established by thorough experimental and theoretical studies~\cite{rodriguez2023beyond}, and the stable and naturally abundant $^{161}$Dy nucleus is expected to possess a large NSM~\cite{Dalton2023enhanced}. Dy-containing piezoelectric solids have already been proposed to probe oscillatory Schiff moments induced by QCD axion dark matter~\cite{Arvanitaki24}.

Here we consider laser-cooled $^{161}$DyO molecules as an intriguing platform for NSM measurements. While a detailed analysis of the laser cooling prospects for DyO is beyond the scope of this work, we briefly explain the potential scheme to motivate the calculation of $W_{\text{NSM}}$. The ground state, $X$, of DyO has $\Omega=8$ and is dominated by the Dy$^{2+}[4f^{9}6s^{1}]$O$^{2-}$ configuration. According to ligand field theory calculations, all excited states expected below $\sim$15,000$\wn$ have $\Omega<8$, so that an optical excitation from $X$ to an $\Omega=9$ state should be electronically closed~\cite{Carette1988ligand}. The three strongest vibronic bands in the range of 400--700~nm excitation wavelengths are all excitations to $\Omega=9$ states with rotational constants closely matching that of $X$~\cite{Kaledin1981electronic}. This suggests the possibility of rapid photon cycling on transitions with favorable vibrational branching ratios. A full laser cooling scheme with $\sim$7 lasers---fewer than used to magneto-optically trap CaOH~\cite{Vilas2022magneto}---may be sufficient to close all significant vibrational and rotational loss channels.

Computed values of $W_{\text{NSM}}$ DyO and DyF (discussed below) are shown in Table~\ref{tab3}. DyO possesses somewhat reduced sensitivity compared to the heavier species considered in this work. For example, $W_{\text{NSM}}$ for DyO is only about 30\% that of the $^{3}\Delta_{1}$ state of ThO, another intrinsically parity-doubled state of a neutral oxide. This is partially explained by the trend that $W_{\text{NSM}}$ scales with atomic number generically faster than $Z^{2}$, and in addition is consistent with the back-polarization of the dysprosium-centered $6s$ electron. Nevertheless, $W_{\text{NSM}}$ is of the same order of magnitude for DyO as for other species of interest.

To elucidate the effect of back-polarization on $W_{\text{NSM}}$ in Dy-containing molecules, we also compute the sensitivity parameter in DyF, which is dominated by a Dy$^{+}[4f^{9}6s^{2}]$F$^{-}$ configuration~\cite{Yamamoto2015electronic,McCarthy1996laser}. As seen in Table~\ref{tab3}, $W_{\text{NSM}}$ is nearly a factor of four smaller in DyF than in DyO. This result is explained by the back-polarization of both $6s$ electrons in DyF. Unlike the case of TlF, the combined effect of the back-polarized $s$ electrons does not overwhelm the effect of the bonding mechanism, so that $W_{\text{NSM}}$ is reduced in magnitude, but does not adopt a positive sign.


\subsection{Discussion on the accuracy of the computational results}

Let us first analyze the convergence of electron-correlation effects.
As shown in Table 3, the triples contributions, 
i.e., the differences between CCSD(T) and CCSD results,
are typically 4-5 times smaller than the singles and doubles contributions. 
This indicates a nice convergence of the CC series in treating electron-correlation effects
on the computed parameters. 
The triples contributions are smaller than 5\% of the total values for all of the systems studied here,
except for the $^1\Sigma^+$ state of ThO, in which the total value exhibits a small absolute magnitude because
of a major cancellation between the contributions from the two competing mechanisms.
The remaining errors in basis sets are no larger than a few percent of the total values,
based on the benchmark studies in the previous section.
We have also investigated the contributions from the correlation of inner-shell electrons
and found that they are also smaller than a few percent of the total values. 
{ {
The comparison of the present X2C-HF results with those from four-component calculations \cite{Hubert22,Marc23}
shows that the errors of the X2C scheme amount to around 1-2 \% of the total values.}}
Therefore, we conclude that the remaining errors in the treatments of electron-correlation, basis-set, { {and relativistic effects}}
are relatively small. 

On the other hand, it is not straightforward to estimate the errors due to the use 
of approximate nuclear charge distributions.
The nuclear charge distributions adopted in the present work
consist of single Gaussian functions.
These may provide reasonable representations for the overall sizes
of the nuclei, but they do not have the correct asymptotic behavior and do not include detailed information about the nuclear structures. 
Therefore, it seems logical to expect the computed results to be qualitatively correct, 
but not quantitatively accurate, despite the fact that 
the treatments of basis-set and electron-correlation effects have nicely converged. 

\section{Summary and Outlook}

We report relativistic coupled-cluster calculations of nuclear Schiff moment (NSM) sensitivity factors
for molecules containing heavy nuclei, 
which are promising avenues for searches for physics beyond the Standard Model 
in the hadronic sector. 
The use of analytic relativistic coupled-cluster gradient techniques greatly expedites the calculations
and enables routine black-box calculations of these parameters. 
Future work will extend the Cholesky-decomposition-based implementation 
for exact two-component coupled-cluster methods \cite{Zhang24} to analytic gradient calculations,
aiming to further improve the computational efficiency by significantly reducing the storage requirement. 
This will enable routine calculations for molecular sensitivity factors 
in medium-sized molecules, 
which might be sufficient to cover the molecular species of interest to the search for the NSM interaction. 
 
We also elucidate two competing chemical mechanisms contributing to the NSM sensitivity factors, 
one from polar chemical bonds and the other from back-polarization of
non-bonding $s$-type orbitals. 
We point out simple, chemically-motivated strategies to engineer molecules with
 large NSM sensitivity factors, by
 maximizing one of the mechanisms and minimizing the other.
TlF illustrated the case in which contributions from back-polarized non-bonding orbitals
 were maximized, while most of the other species considered relied on highly polar chemical bonds
and minimal contributions from back-polarization.
While this chemical intuition is a helpful guide, it is in general still necessary to perform first-principle calculations to determine NSM sensitivity factors quantitatively. 

\section*{Acknowledgments}

Z. L. is grateful to Svetlana Kotochigova for stimulating the investigation into Dy-containing molecules. L. C. is grateful to Leonid Skripnikov for communicating his DC-HF calculations and graciously granting the permission to quote the DC-HF results in the present manuscript.

The theoretical and computational work at the Johns Hopkins University 
was supported by the
National Science Foundation, under Grant No. PHY-2309253.
The computations were carried out at
Advanced Research Computing at Hopkins (ARCH) core
facility (rockfish.jhu.edu), which is supported by the NSF 
under Grant OAC-1920103.  
TRIUMF is supported by the Natural Sciences and Engineering Council of Canada (NSERC). TRIUMF receives federal funding via a contribution agreement with the National Research Council of Canada.
The work at University of New South Wales was supported by the Australian Research Council Grants No. DP230101058 and DP200100150.
The work at Harvard University has been supported by 	
the Center for Ultracold Atoms (CUA), an NSF Physics Frontier Center. 
B. A. gratefully acknowledges Williams College for startup funds to support work at Williams. Work at Caltech was supported by the
National Science Foundation under Grant Nos. PHY-1847550 (CAREER) and PHY-2309361.  A. M. J. acknowledges the support of the U. S. Department of Energy (DE-SC0022034). The work at University of Chicago was supported by the U. S. Department of Energy (DE-SC0024667) and by the Gordon and Betty Moore Foundation (Grant No. 12330).

\section*{Supplementary Material}

The structures, basis sets, and the number of frozen core orbitals used in the calculations. 

\section*{Declaration of interest statement}

The authors report that there are no competing interests to declare.

\newpage  

\clearpage

\bibliography{reference0,reference1,ref-laser-cooling,referenceKB,referencesDy,referencesMQMNSMintro,reference-nrh}






\end{document}